\documentclass[preprint,showkeys,aps,prb]{revtex4}
\usepackage[dvips]{graphicx}
\begin{document}

\title{
Bit-Reversible Version of Milne's Fourth-Order Time-Reversible Integrator for Molecular Dynamics
}

\author{
William Graham Hoover and Carol Griswold Hoover \\
Ruby Valley Research Institute                  \\
Highway Contract 60, Box 601                    \\
Ruby Valley, Nevada 89833                       \\
}

\date{\today}


\vspace{0.1cm}

\begin{abstract}
We point out that two of Milne's fourth-order integrators are well-suited to bit-reversible simulations.
The fourth-order method improves on the accuracy of Levesque and Verlet's algorithm and simplifies the
definition of the velocity $v$ and energy $e = (q^2 + v^2)/2$ .  ( We use this one-dimensional oscillator
problem as an illustration throughout this paper ). Milne's integrator is particularly useful
for the analysis of Lyapunov ( exponential ) instability in dynamical systems, including manybody
molecular dynamics.  We include the details necessary to the implementation of Milne's Algorithms.
\end{abstract}

\maketitle

\section{Introduction}

William Milne's 1949 work {\it Numerical Calculus}\cite{b1} was republished by the Princeton University
Press in 2015. The book is a particularly valuable source of clear and direct numerical methods. Research
workers in statistical mechanics, molecular dynamics, and dynamical systems will find his approach to
what is our own research interest, solving and analyzing differential equations for chaotic systems small
and large, reliable and useful.  Writing about a decade prior to the computer revolution Milne had no
particular interest in ``reversible computing'' and the ``bit-reversible'' algorithms which make it
possible to extend sequences of coordinates forward and backward in time stably and reversibly {\it ad
infinitum}.  Nevertheless his work is directly applicable to such finite-difference applications.

In 1993 Dominique Levesque and Loup Verlet used an {\it integer} algorithm to solve problems in Newtonian
mechanics with perfect time reversibility\cite{b2}. Loup had popularized St{\o}rmer and Newton's
Leapfrog Algorithm a quarter century earlier, in the early days of molecular dynamics\cite{b3},
$$
q_{t+dt} - 2q_t + q_{t-dt} = a_t(dt)^2 \ .
$$
``Verlet's algorithm'' appears on page 140 of Reference 1.
If the righthand side of this finite-difference algorithm is truncated to an integer the resulting acceleration
is precisely the same ( to the very last computational ``bit'' ) in either direction of time.  Because this
algorithm conserves phase volume when written in a ``symplectic'' centered-difference form :
$$
q_{t+(dt/2)} = q_t + v_t(dt/2) \ ; \ v_{t+dt} = v_t + a_{t+(dt/2)}dt \ ; \ q_{t+dt} = q_{t+(dt/2)} + v_{t+dt}(dt/2) \ ,
$$
there is no tendency for energy drift. The errors in the velocity and energy in the leapfrog
algorithm are unnecessarily large, so that two of Milne's algorithms ( both of them also on page 140 of
Reference 1 ) can provide better accuracy for longer runs :
$$
q_{t+2dt} - q_{t+dt} - q_{t-dt} + q_{t-2dt} = [ \ 5a_{t+dt} + 2a_t + 5a_{t-dt} \ ](dt^2/4) \ .
$$
The error, $\simeq(17/240)dt^6$, is quite tolerable relative to the St{\o}rmer error, $\simeq (1/12)dt^4$.
Milne also gives an even better corrector formula with an error $\simeq(-1/240)dt^6$ .
$$
q_{t+2dt} - 2q_{t+dt} + q_{t} = [ \ a_{t+2dt} + 10a_{t+dt} + a_{t} \ ](dt^2/12) \ .
$$

\section{Applications}

For several years now\cite{b4,b5} we have been exploring the differences in Lyapunov spectra forward and
backward in time in order to get insight into the Second Law of Thermodynamics.  The fractal structures
which arise in nonequilibrium deterministic and time-reversible steady-state problems provide explanations
to both Loschmidt's Reversibility paradox and Zerm\'elo's Recurrence paradox\cite{b6}.  Levesque and
Verlet's integer algorithm has proved to be a useful tool in these studies despite its relatively coarse
description of particle trajectories.

Integer algorithms are also useful in studies of the effects of finite precision ( single, double,
quadruple, ... ) on phase-space distributions generated by flows and maps\cite{b7,b8}. Mauricio
Romero-Bastida\cite{b9} suggested the use of the integer-based leapfrog algorithm for generating a reversible
reference trajectory of arbitrary length in his studies of the ``covariant'' Lyapunov exponents.  In
2013 we were able to see a qualitative difference between the ``important particles'' ( those making
above-average contributions to the Lyapunov instability ) forward and backward
in time in the example inelastic-collision problem of {\bf Figure 1}\cite{b10}.  Continuing progress in
low-cost computation caused us to revisit these problems in connection with a lecture course delivered
at Kharagpur's Indian Institute of Technology in December 2016\cite{b11}.  We were very pleased to find
that Milne's work offers an improvement in the precision and accuracy of these Lyapunov studies and
believe that others will find this approach useful to their own work.  Although these improvements are
not at all ``new'' we do expect that this work will accelerate progress in understanding the time-reversible
simulation of irreversible processes.

{\bf Figure 1} illustrates the important particles ( those making above-average contributions to the
largest Lyapunov exponent ) forward and backward in time for the collision of two 400-particle balls.
The 162 important particles forward in time are those blacked in along the interface between the balls while
the 120 important particles backward in time are those blacked in in the necking regions where the plastic
strain is greatest as the balls are separating.  This simulation employed the Levesque-Verlet algorithm
for the reference trajectory and a Runge-Kutta fourth-order algorithm for the two satellite trajectories
( one forward and another backward ) as is described in Reference 10.

\begin{figure}[h]
\centerline{\includegraphics[width=2.0in,angle=-90,bb=20 105 592 690]{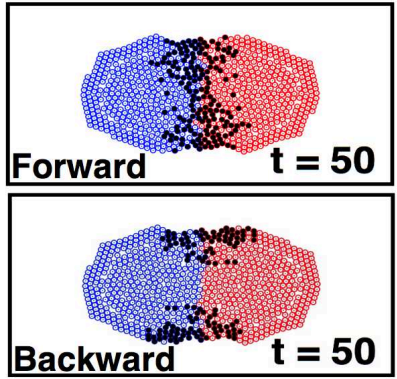}}
\caption{
Shows the important particles evaluated both forward and backward in
time at a time of 50.  The important particles are blacked in in the figure. The two
colliding balls are inverted images of each other.  We used the Levesque-Verlet
algorithm to reverse the motion at time 100 and analyze the trajectory ``backward''
in time to the pre-collision state at time 0.  The bit-reversible algorithm makes it unnecessary
to store a forward trajectory in order to process it backward. Notice in the
reversed trajectory, where the coalesced balls are beginning to ``neck'' in order to
separate, that the important particles are on the periphery, where the strain is greatest,
rather than at the interface.  Although the motion reverses perfectly the stability does
not.  It is remarkable that the two directions of time differ so much ( 162 forward {\it versus}
120 backward) in this important-particle measure of chaos. For additional snapshots in this
series see Reference 10.
}
\label{Figure 1}
\end{figure}

\section{Numerical Implementation of Milne's Algorithms}

To illustrate the application of Milne's Algorithms  we consider an integer version of the
simpler of his two fourth-order algorithms.  We describe a harmonic oscillator with $\ddot q = -q$ .
The preliminaries, which we give below, provide integer forms for five previous coordinates and the
corresponding contributions to the acceleration, all of them multiplied by $10^{15}$. We select an
example timestep of $(\pi/50)$ in the Fortran instructions so that an oscillator period corresponds
to one hundred timesteps.

We carried out two kinds of tests for the Milne integrator, reversibility, confirming that reversing the four
prereversal coordinates exactly reverses the sequence of integers back to the initial value of $10^{15}$.
It is easy to show that the algorithm is exactly reversible in this way.  Stability can be confirmed
by solving for the dependence of the oscillation frequency on the timestep. Numerical work consistent
with the linear analysis for the oscillator ( given in more detail in a Postscript ) shows that
the dependence of the phase shift is quartic in the timestep for the range $0 < dt < 1$.  A direct
simulation of the integer version of the algorithm, using two billion timesteps with $dt = 0.2$, showed no tendency 
toward damping or instability.  Similar results can be obtained by solving the floating-point version
of the problem, where precise reversibility has to be abandoned ( because roundoff error will spoil it ).
These results establish that the Milne algorithm is both reversible and stable for the oscillator. We recommend
it to our colleagues for their use.

The implementation of the algorithm is to some extent hardware dependent. On our various Mac computers
using the free gnu compiler we had no trouble using 16-byte integers, giving roughly 15 digits for
arithmetical operations.  The following extract from the setup of the computation generates the inital
data ( in this case four points from a cosine curve ) as well as the three integers, proportional to
$dt^2\times 10^{15}$, and  needed for the accelerations.  On the following page we summarize the time-stepping loop
where the three accelerations are expressed as integers $\{ {\tt IAP,IA0,IAM } \}$.  We include at the end an
indication of the coordinate reversal procedure needed to integrate backward.

\begin{verbatim}
      IMPLICIT DOUBLE PRECISION (A-H,O-Z)
      INTEGER*16 IQMM,IQM,IQ0,IQP,IQPP  ! Contiguous integer coordinates
      INTEGER*16 IAP,IA0,IAM            ! Ingredients of the accelerations
      INTEGER*16 I1,I2,I3,I4            ! Storage for coordinate reversal
      ITMAX = 100
      TWOPI = 2.D0*3.141592653589793D0
      DT = TWOPI/ITMAX
      QMM  =  DCOS(-2.D0*DT)
      QM   =  DCOS(-1.D0*DT)
      Q0   =  DCOS( 0.D0*DT)
      QP   =  DCOS(+1.D0*DT)
CONVERT COORDINATES TO 15-DIGIT INTEGERS
      IQMM   = QMM*(10.D0**15)  
      IQM    = QM *(10.D0**15)  
      IQ0    = Q0 *(10.D0**15)   
      IQP    = QP *(10.D0**15)          ! Finish of the preliminaries
 
\end{verbatim}

\pagebreak
Here follows a bare-bones evolution loop for the integer coordinates. After {\tt ITMAX} iterations the
coordinate reversal steps make it possible to return precisely to, and beyond, the beginning. In the
event that the velocities are to be calculated from Milne's fifth-order interpolation ( which is one
order of overkill ) it is necessary to
compute the integer coordinates to include {\tt IQPPP} and {\tt IQMMM}, getting {\tt IQPPP} from
the ``step'' {\tt IQPPP = IQPP + IQ0 - IQM - (IAPP + IAP + IA0)}.
\begin{verbatim}
      DO IT = 1,ITMAX
      TIME = IT*DT
COMPUTE INGREDIENTS OF THREE  ACCELERATIONS
      IAP  = 0.25D00*DT*DT*(5.D0*IQP)
      IA0  = 0.25D00*DT*DT*(2.D0*IQ0)
      IAM  = 0.25D00*DT*DT*(5.D0*IQM)       
COORDINATE UPDATES FOR FIVE SUCCESSIVE TIMES
      IQPP = IQP + IQM - IQMM - (IAP + IA0 + IAM)  ! This is the "step".
      IQMM = IQM
      IQM  = IQ0
      IQ0  = IQP
      IQP  = IQPP
      END DO
COORDINATE REVERSAL
      I1 = IQMM 
      I2 = IQM
      I3 = IQ0
      I4 = IQP
      IQMM  = I4
      IQM   = I3
      IQ0   = I2
      IQP   = I1
      DO IT = 1,IMAX
CONTINUE REVERSAL WITH SAME STATEMENTS AS IN THE FORWARD LOOP
      END DO
\end{verbatim}
\pagebreak
There is no difficulty in computing an accurate velocity with Milne's page 99 formula using
six centered coordinates. This fifth-order interpolation gives not only good velocities,
but also an accurate energy.  Accurate values of these phase variables are a real advantage
of the Milne algorithm over that of Levesque and Verlet.

\section{Postscript on the Stability of Milne's Algorithm}

The stability analysis for Milne's algorithm is straightforward. If we substitute
the trial solution $q \propto e^{i\omega t}$ into the wholly linear algorithm the result is :
$$
\cos(2\omega dt) - \cos(\omega dt) + (dt^2/4)[ \ 5\cos(\omega dt) + 1 \ ] = 0 \ .
$$
This simplifies to a quadratic equation in $\cos(\omega dt)$ :
$$                                                                                                                                                                 
2C^2 + [ \ (5dt^2/4) - 1 \ ]C  + (dt^2/4) - 1  = 0 \ {\rm where} \ C \equiv \cos(\omega dt) .
$$
{\bf Figure 2} shows that the dependence of the frequency error on the timestep is quartic,
$(1 - \omega) \propto dt^4$, confirming the stability of the algorithm.

\begin{figure}[h]
\centerline{\includegraphics[width=3.0in,angle=-90,bb=75 61 543 717]{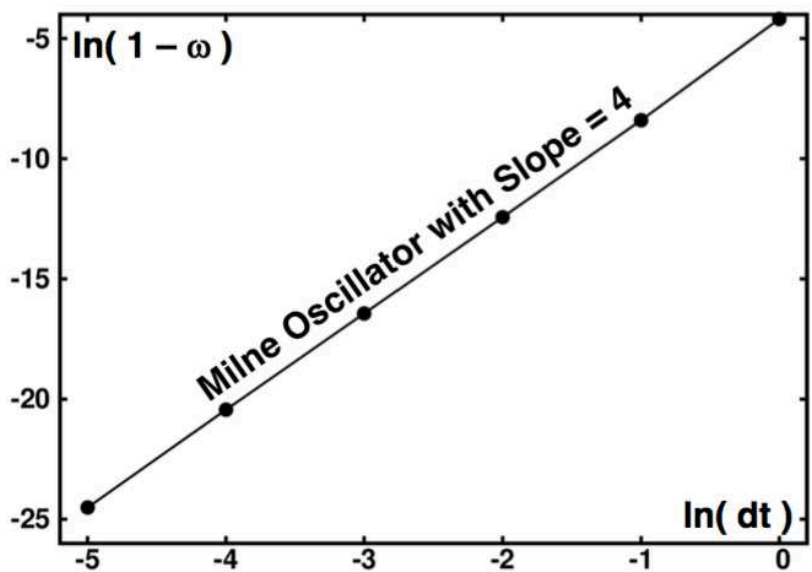}}
\caption{
Analysis of the oscillator motion equation confirms the stability of Milne's method.
}
\label{Figure 2}
\end{figure}

\section{Acknowledgement}  The interest and support of Harald Posch ( Universit\"at Wien ),
Clint Sprott ( University of Wisconsin-Madison ), and Karl Travis ( University of Sheffield )
are gratefully acknowledged.

\end{document}